\documentclass[epj]{webofc}
\usepackage[varg]{txfonts}   
\woctitle{MENU 2013}
\begin{document}
\title{Generalized parton distributions from neutrino experiments: twist-three effects}

\author{B. Z. Kopeliovich\inst{1}\fnsep\thanks{\email{Marat.Siddikov@usm.cl}}\and
Iv\'an Schmidt\inst{1}\fnsep\thanks{\email{Boris.Kopeliovich@usm.cl}} \and
        M. Siddikov\inst{1}\fnsep\thanks{\email{Ivan.Schmidt@usm.cl}} 
}

\institute{Departamento de F\'isica, Universidad T\'ecnica Federico Santa Mar\'ia,\\
 y Centro Cient\'ifico - Tecnológico de Valpara\'iso, Avda. Espa\~na 1680, Valpara\'iso,
Chile}

\abstract{%
  We study the twist-3 corrections to the neutrino induced deeply
virtual meson production due to the chiral odd transversity Generalized Parton Distribution (GPD). We found
that in contrast to pion \emph{electro}production, in neutrino-induced processes
these corrections are small. This occurs due to large contribution
of unpolarized GPDs $H,\, E$ to the twist-2 amplitude in neutrinoproduction.
Our results are important for analyses of the pion and kaon production in the \textsc{Minerva} experiment at FERMILAB. 
}
\maketitle
\section{Introduction}

The generalized parton distributions (GPDs) are important phenomenological objects parametrizing the nonperturbative
structure of the target. In the Bjorken kinematics, in which the collinear factorization is applicable~\cite{Ji:1998xh,Collins:1998be},
the cross-sections for a wide class of processes can be expressed in terms of GPDs.
Nowadays all available information on GPDs is provided by the electron-proton and
positron-proton experiments at Jefferson Lab and HERA, particularly by measurements of  deeply virtual Compton scattering (DVCS) and deeply virtual meson
production (DVMP)~\cite{Ji:1998xh,Collins:1998be,Mueller:1998fv,Ji:1998pc,Radyushkin:1996nd,Radyushkin:1997ki,Collins:1996fb,Brodsky:1994kf,Diehl:2000xz,Belitsky:2001ns,Belitsky:2005qn}.
The future constraints on GPD parametrizations will come from upgraded CLAS12 at JLAB~\cite{Kubarovsky:2011zz} and muon-induced DVCS and DVMP measurements at
COMPASS~\cite{Ferrero:2011zz,Kouznetsov:2011zza}. In a more distant future, EIC~\cite{Accardi:2012qut} 
and LHeC~\cite{AbelleiraFernandez:2012cc} machines could also contribute to a better understanding of the GPDs.

In addition to the virtual photon-mediated processes, extra constraints on GPD parametrizations can be inferred from deeply virtual neutrinoproduction of the pseudo-Goldstone mesons ($\pi,\, K,\,\eta$), as we recently suggested in~\cite{Kopeliovich:2012dr}. The $\nu$DVMP process is complementary to the $e$DVMP. Due to the $V-A$ structure of the charged current, in $\nu$DVMP one can access simultaneously the unpolarized GPDs, $H,\, E$, and the helicity flip GPDs, $\tilde{H}$ and $\tilde{E}$.
The produced Goldstone mesons due to chiral symmetry have very close characteristics and in this way act as natural probes for the flavor content, enabling us to extract the full flavour structure of the GPDs. Experimentally neutrino-induced DVMP could be studied with the high-intensity~\textsc{NuMI} beam at Fermilab, which recently switched to the middle-energy (ME) regime with the mean neutrino energy of about 6~GeV. Potentially the energy can reach  20 GeV without essential loss of luminosity. 

It is worth reminding that the cross-sections were evaluated  in~\cite{Kopeliovich:2012dr}
in the leading twist approximation, and for a correct extraction of the
GPDs at the energies of MINERvA in ME regime, an estimate of the higher
twist effects is required. The first twist-3 correction arises due to
contribution of the transversely polarized intermediate virtual bosons and is
controlled by convolution of the poorly known transversity GPDs $H_{T},\, E_{T},\,\tilde{H}_{T},\,\tilde{E}_{T}$
and twist-3 DAs of pion. While this correction vanishes
at asymptotically large $Q^{2}$, in \emph{electro}production at moderate $Q^{2}$ 
it gives a sizable contribution, as was confirmed
by CLAS collaboration~\cite{Kubarovsky:2011zz}. In the case of neutrino-production
the situation is different because
there is an additional and numerically dominant contribution of the
unpolarized GPDs $H,\, E$ to the leading twist amplitude due to the $V-A$ structure of the weak currents. In what follows we analyze the relative magnitude of the twist-3 contributions to the neutrino-production of pions and demonstrate that they are indeed
small. In this respect we are different from~\cite{Goldstein:2009in}, where
the contribution of the chiral odd GPDs was assumed to be numerically
dominant.

\section{Cross-section of the $\nu$DVMP process}

\label{sec:DVMP_Xsec}The cross-section of the Goldstone mesons production
in  neutrino-hadron collisions has the form
\begin{align}
\frac{d\sigma}{dt\, dx_{B}dQ^{2}d\phi} & =\epsilon\frac{d\sigma_{L}}{dt\, dx_{B}dQ^{2}d\phi}+\frac{d\sigma_{T}}{dt\, dx_{B}dQ^{2}d\phi}+\sqrt{\epsilon(1+\epsilon)}\cos\phi\frac{d\sigma_{LT}}{dt\, dx_{B}dQ^{2}d\phi}\label{eq:sigma_def}\\
 & +\epsilon\cos2\phi\frac{d\sigma_{TT}}{dt\, dx_{B}dQ^{2}d\phi}+\sqrt{\epsilon(1+\epsilon)}\sin\phi\frac{d\sigma_{L'T}}{dt\, dx_{B}dQ^{2}d\phi}+\epsilon\sin2\phi\frac{d\sigma_{T'T}}{dt\, dx_{B}dQ^{2}d\phi},\nonumber 
\end{align}
where $t=\left(p_{2}-p_{1}\right)^{2}$ is the momentum transfer to
baryon, $Q^{2}=-q^{2}$ is the virtuality of the charged boson, $x_{B}=Q^{2}/(2p\cdot q)$
is Bjorken $x$, $\phi$ is the angle between the lepton and meson
production scattering planes, and we introduced shorthand notations
\[
\epsilon=\frac{1-y-\frac{\gamma^{2}y^{2}}{4}}{1-y+\frac{y^{2}}{2}+\frac{\gamma^{2}y^{2}}{4}},\quad\gamma=\frac{2\, m_{N}x_{B}}{Q},\quad y=\frac{Q^{2}}{sx_{B}}.
\]
 In the asymptotic Bjorken limit the cross-section is dominated by
the first angular independent term $\epsilon\, d\sigma_{L}/dt\, dx_{B}dQ^{2}d\phi$
which was studied in our previous paper~\cite{Kopeliovich:2012dr}
and is a straightforward extension of the electroproduction of pions
studied in~\cite{Vanderhaeghen:1998uc,Mankiewicz:1998kg,Goloskokov:2006hr,Goloskokov:2007nt,Goloskokov:2008ib,Goloskokov:2011rd,Goldstein:2012az}.
As we will see below the twist-3 corrections are small, for this reason
it is convenient to normalize all the cross-sections in~(\ref{eq:sigma_def})
to this term,
\begin{equation}
\frac{d\sigma}{dt\, dx_{B}dQ^{2}d\phi}=\epsilon\frac{d\sigma_{L}}{dt\, dx_{B}dQ^{2}d\phi}\sum_{n}\left(c_{n}\cos n\phi+s_{n}\sin n\phi\right)\label{eq:harmonicsDefinition}
\end{equation}
and discuss the higher-twist effects in terms of harmonics $c_{n},\, s_{n}$. Here we omit the details of evaluation and refer the reader to~\cite{Kopeliovich:2014pea}.
The most important harmonics is $c_{0}$, because its deviation from unity affects the extraction of GPDs in the leading twist approximation, which requires an experimentally challenging Rosenbluth separation with varying energy neutrino beam.
All the other harmonics generate nontrivial angular dependence and can be easily separated from the leading-twist contribution. For example, the
angle-integrated cross-section $d\sigma/d\ln x_{B}dt\, dQ^{2}$ is not sensitive to those harmonics at all.

For numerical estimates we used the Kroll-Goloskokov parametrization of GPDs~\cite{Goloskokov:2006hr,Goloskokov:2007nt,Goloskokov:2008ib,Goloskokov:2009ia}.

\begin{figure}
\includegraphics[scale=0.25]{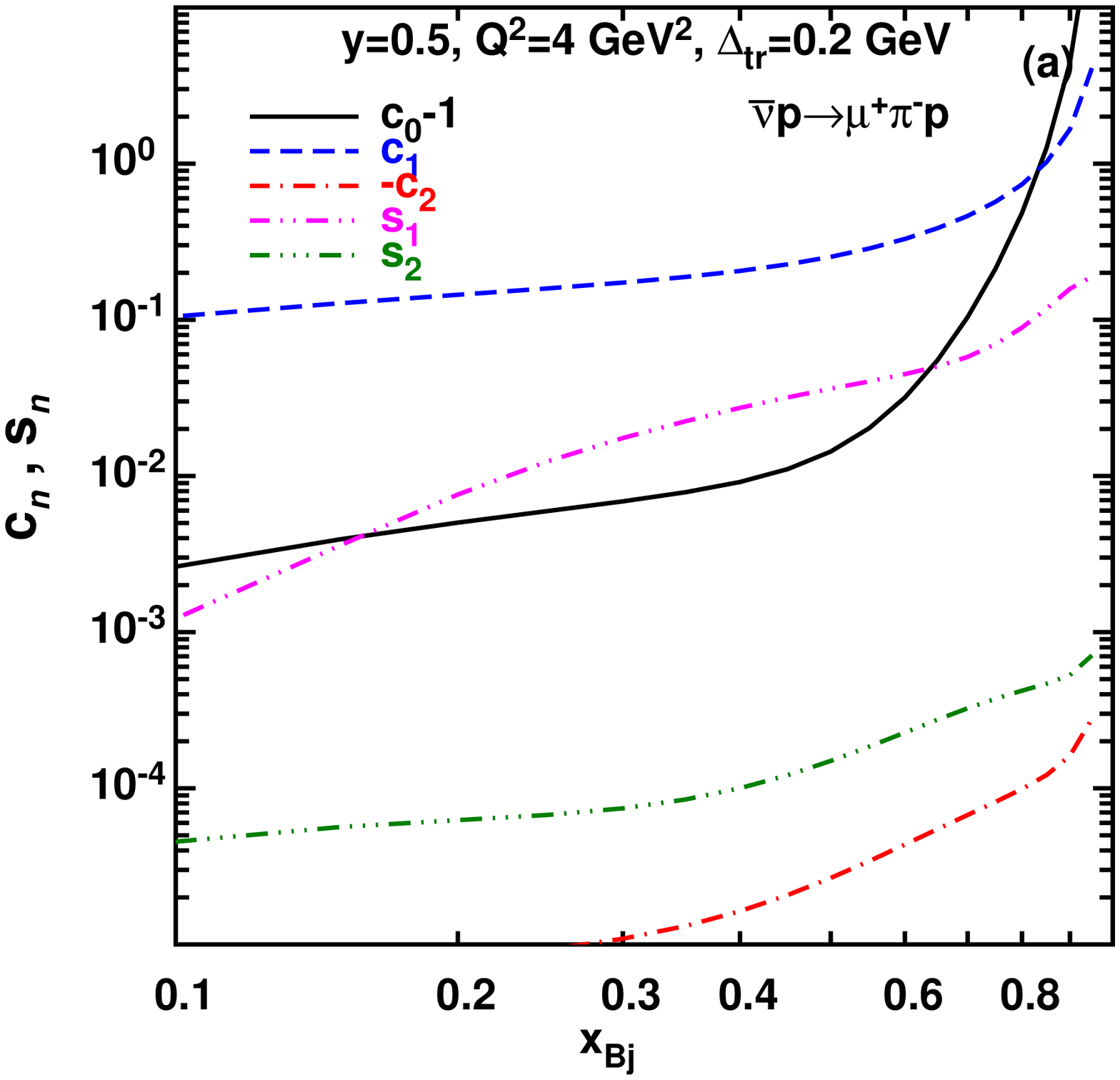}\includegraphics[scale=0.25]{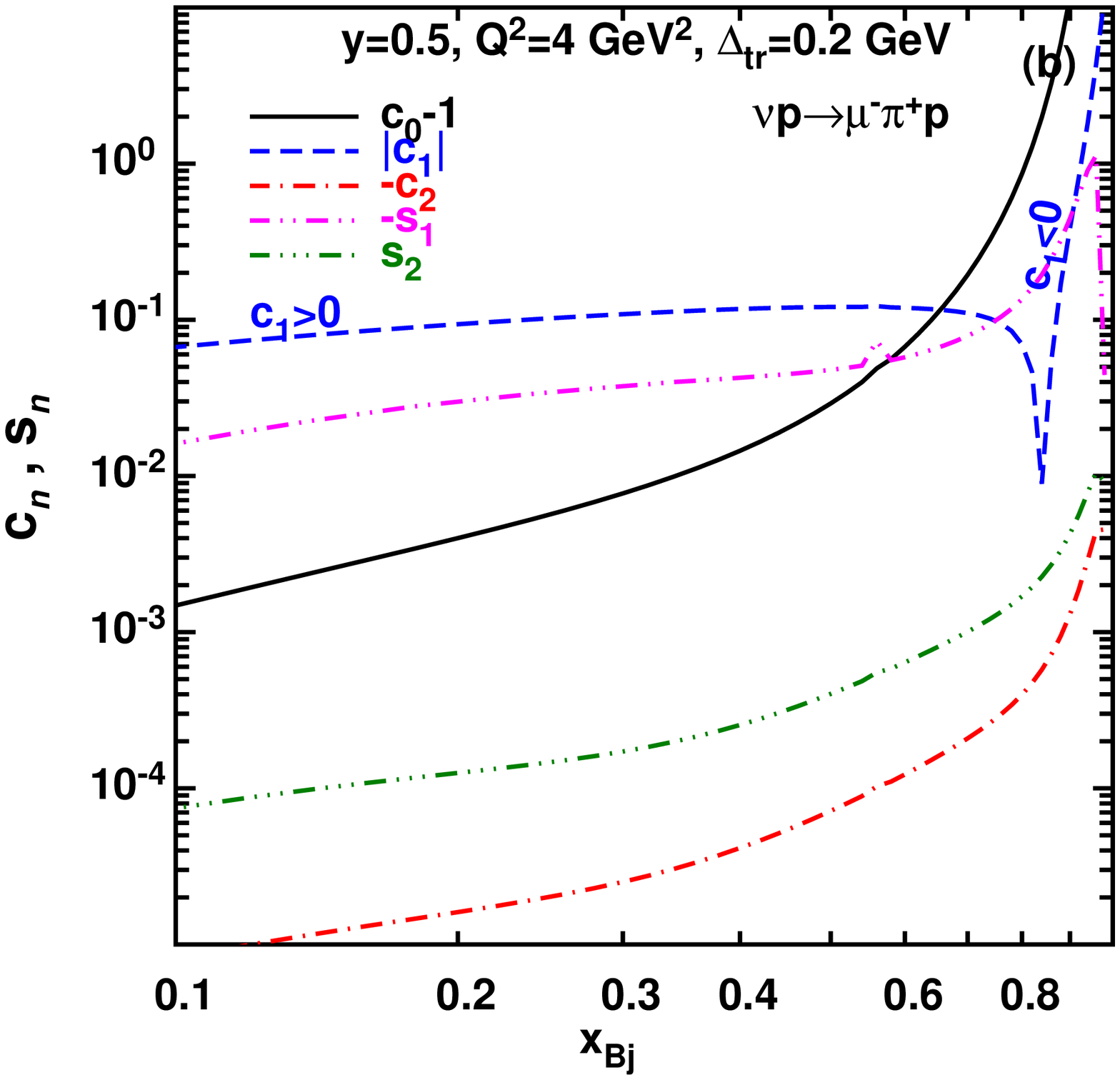}\includegraphics[scale=0.25]{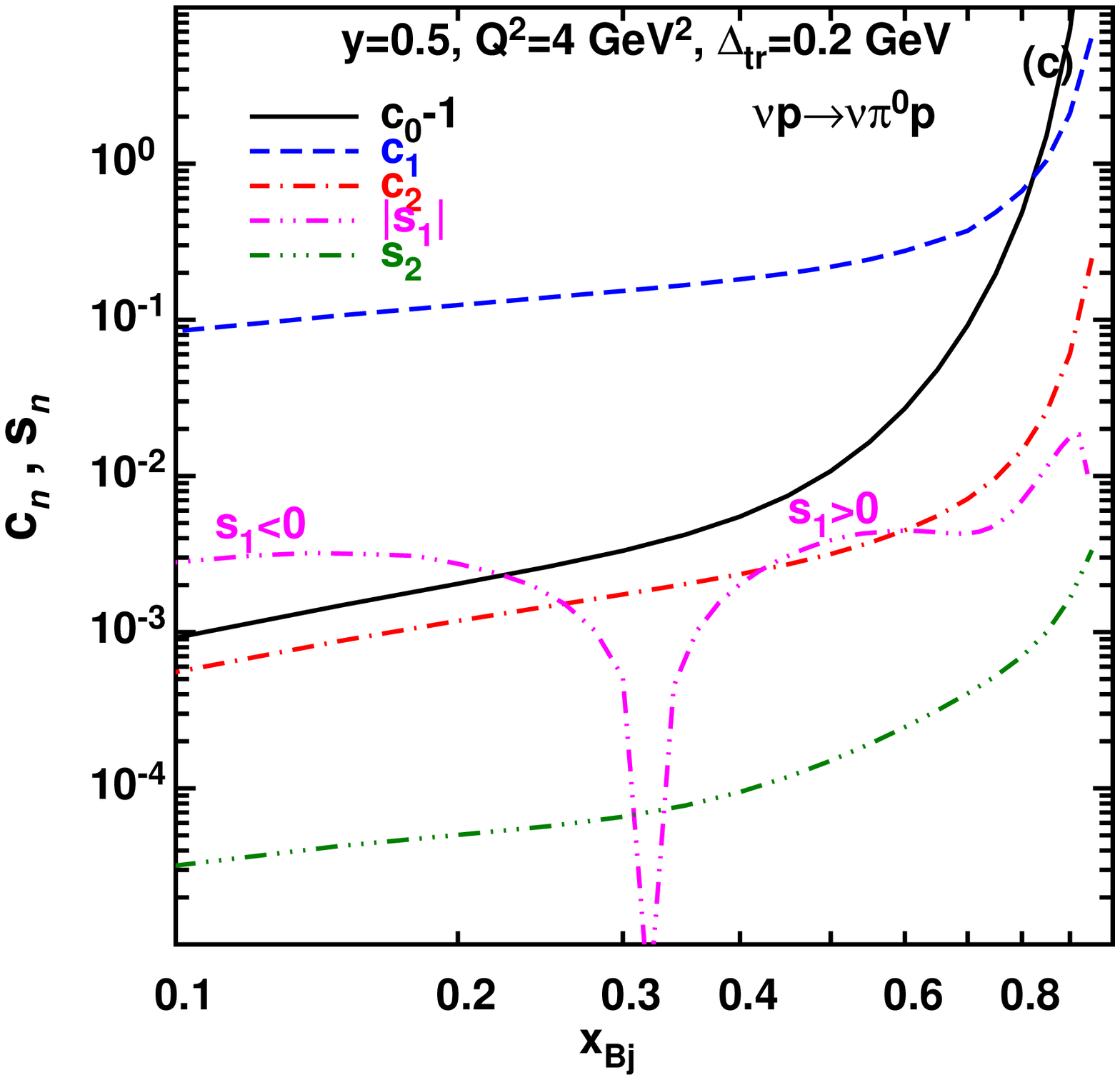}\\
\caption{\label{fig:DVMP-pions}Pion production on nucleons. See~(\ref{eq:harmonicsDefinition}) for definition of harmonics $c_n,\,s_n$.}
\end{figure}

In Figure~\ref{fig:DVMP-pions} we show the harmonics $c_{n},\, s_{n}$
for some processes. At $x_B\lesssim0.5$, where the cross-section is
the largest, the harmonics are small and do not exceed few per cent.
The largest twist-3 contribution is due to the $c_{1}$ harmonics,
which can reach up to twenty percents. This is different from the electroproduction
experiments, where $c_{1}$ ($\sim\sigma_{LT}$) is very small: due to parity
nonconservation in weak interactions we have for the interference
term $\sigma_{0+}\not=\sigma_{0-}$. A positive value of $c_{1}$
for most processes implies that pion production correlates
with the direction of the produced muon (scattered neutrino) in the case of
CC (NC) mediated processes. The interference term also yields a relatively
large harmonics $s_{1}$ which appears due to the interference of the
vector and axial vector contributions. 

In the region of $x\gtrsim0.5$ all the harmonics increase, but the
cross-sections for both the leading twist and subleading twist results
are suppressed there due to increase of $|t_{min}|$ and are hardly
accessible with ongoing and forthcoming experiments.

Similar results can be obtained for the processes with change of the baryon state and for processes with strangeness production.

Notice that similar angular harmonics can be generated by interference of the leading
twist result with the electromagnetic corrections~\cite{Kopeliovich:2013ae}.
At moderate virtualities of the order a few ${\rm GeV}^{2}$ this mechanism
also gives small harmonics (of the order few per cent), however those
corrections grow rapidly as a function of $Q^{2}$, and already at
$Q^{2}\sim100\,{\rm GeV^{2}}$ electromagnetic mechanism becomes dominant.

To summarize, we conclude that deeply virtual production of pions and
kaons on protons and neutrons by neutrinos with typical values of $Q^{2}$ of the
order few ${\rm GeV}^{2}$ provides a theoretically clean probe for the GPDs, with
various corrections of the order of few per cent. Our results are relevant
for analysis of the pion and kaon production in the \textsc{Minerva
}experiment at FERMILAB as well as for the planned~Muon Collider/Neutrino
Factory~\cite{Gallardo:1996aa,Ankenbrandt:1999as,Alsharoa:2002wu}.
An optimal target for study of the GPDs could be hydrogen
or deuterium. For other targets there is an additional uncertainty
due to the nuclear effects which will be addressed elsewhere.

\section*{Acknowledgments}

This work was supported in part by Fondecyt (Chile) grants No. 1130543,
1100287 and 1120920.

\appendix


\begin{thebibliography}{10}
   \bibitem{Ji:1998xh} X.~D.~Ji and J.~Osborne, Phys.\ Rev.\ D
   \textbf{58} (1998) 094018 [arXiv:hep-ph/9801260].
   
   \bibitem{Collins:1998be} J.~C.~Collins and A.~Freund, Phys.\ Rev.\ D
   \textbf{59}, 074009 (1999).
   
   \bibitem{Mueller:1998fv} D.~Mueller, D.~Robaschik, B.~Geyer, F.~M.~Dittes
   and J.~Horejsi, Fortsch.\ Phys.\ \textbf{42}, 101 (1994) [arXiv:hep-ph/9812448].
   
   
   \bibitem{Ji:1998pc} X.~D.~Ji, J.\ Phys.\ G \textbf{24}, 1181
   (1998) [arXiv:hep-ph/9807358].
   
   \bibitem{Radyushkin:1996nd} A.~V.~Radyushkin, Phys.\ Lett.\ B
   \textbf{380}, 417 (1996) [arXiv:hep-ph/9604317].
   
   \bibitem{Radyushkin:1997ki} A.~V.~Radyushkin, Phys.\ Rev.\ D
   \textbf{56}, 5524 (1997).
   
   
   \bibitem{Collins:1996fb} J.~C.~Collins, L.~Frankfurt and M.~Strikman,
   Phys.\ Rev.\ D \textbf{56}, 2982 (1997).
   
   \bibitem{Brodsky:1994kf} S.~J.~Brodsky, L.~Frankfurt, J.~F.~Gunion,
   A.~H.~Mueller and M.~Strikman, Phys.\ Rev.\ D \textbf{50}, 3134
   (1994).
   
   
   \bibitem{Diehl:2000xz} M.~Diehl, T.~Feldmann, R.~Jakob and P.~Kroll,
   Nucl.\ Phys.\ B \textbf{596}, 33 (2001) [Erratum-ibid.\ B \textbf{605},
   647 (2001)] [arXiv:hep-ph/0009255].
   
   \bibitem{Belitsky:2001ns} A.~V.~Belitsky, D.~Mueller and A.~Kirchner,
   Nucl.\ Phys.\ B \textbf{629}, 323 (2002) [arXiv:hep-ph/0112108].
   
   \bibitem{Belitsky:2005qn} A.~V.~Belitsky and A.~V.~Radyushkin,
   Phys.\ Rept.\ \textbf{418}, 1 (2005) [arXiv:hep-ph/0504030].
   
   \bibitem{Kubarovsky:2011zz}V.~Kubarovsky [CLAS Collaboration],
   Nucl.~Phys.~Proc.~Suppl.~\textbf{219-220}, 118 (2011).
   
 \bibitem{Ferrero:2011zz} 
   A.~Ferrero [COMPASS Collaboration],
   J.\ Phys.\ Conf.\ Ser.\  {\bf 295}, 012039 (2011).
 \bibitem{Kouznetsov:2011zza} 
   O.~Kouznetsov [COMPASS Collaboration],
   AIP Conf.\ Proc.\  {\bf 1350}, 69 (2011).
 \bibitem{Accardi:2012qut} 
   A.~Accardi, J.~L.~Albacete, M.~Anselmino, N.~Armesto, E.~C.~Aschenauer, A.~Bacchetta, D.~Boer and W.~Brooks {\it et al.},
   arXiv:1212.1701 [nucl-ex].
 \bibitem{AbelleiraFernandez:2012cc} 
   J.~L.~Abelleira Fernandez {\it et al.}  [LHeC Study Group Collaboration],
   J.\ Phys.\ G {\bf 39}, 075001 (2012)
   [arXiv:1206.2913 [physics.acc-ph]].
  
   \bibitem{Kopeliovich:2012dr}B.~Z.~Kopeliovich, I.~Schmidt and
   M.~Siddikov, Phys. Rev. D \textbf{86} (2012), 113018 [arXiv:1210.4825
   [hep-ph]].
   
   \bibitem{Goldstein:2009in}G.~R.~Goldstein, O.~G.~Hernandez, S.~Liuti
   and T.~McAskill, AIP Conf.~Proc.~\textbf{1222}, 248 (2010) [arXiv:0911.0455
   [hep-ph]].
   
   \bibitem{Vanderhaeghen:1998uc}M.~Vanderhaeghen, P.~A.~M.~Guichon
   and M.~Guidal, 
    Phys.~Rev.~Lett.~\textbf{80}, 5064 (1998).
   
   \bibitem{Mankiewicz:1998kg}L.~Mankiewicz, G.~Piller and A.~Radyushkin,
   \textbf{10}, 307 (1999) [hep-ph/9812467].
   
   \bibitem{Goloskokov:2006hr}S.~V.~Goloskokov and P.~Kroll, 
    Eur.~Phys.~J.~C \textbf{50}, 829 (2007) [hep-ph/0611290].
   
   \bibitem{Goloskokov:2007nt}S.~V.~Goloskokov and P.~Kroll, 
    Eur.~Phys.~J.~C \textbf{53}, 367 (2008) [arXiv:0708.3569 [hep-ph]].
   
   \bibitem{Goloskokov:2008ib}S.~V. Goloskokov and P.~Kroll, Eur.~Phys.~J.~C
   \textbf{59} (2009) 809 [arXiv:0809.4126 [hep-ph]].
   
   \bibitem{Goloskokov:2011rd}S.~V.~Goloskokov and P.~Kroll, Eur.~Phys.~J.~A
   \textbf{47}, 112 (2011) [arXiv:1106.4897 [hep-ph]].
   
   \bibitem{Goldstein:2012az}G.~R.~Goldstein, J.~O.~G.~Hernandez
   and S.~Liuti, arXiv:1201.6088 [hep-ph].
   
  \bibitem{Kopeliovich:2014pea} 
    B.~Z.~Kopeliovich, I.~Schmidt and M.~Siddikov,
    arXiv:1401.1547 [hep-ph].
   
  
   \bibitem{Goloskokov:2009ia}S.~V.~Goloskokov and P.~Kroll, Eur.~Phys.~J.~C
   \textbf{65}, 137 (2010) [arXiv:0906.0460 [hep-ph]].
   
   \bibitem{Kopeliovich:2013ae} B.~Z.~Kopeliovich, I.~Schmidt and
   M.~Siddikov, Phys.\ Rev.\ D \textbf{87}, 033008 (2013) [arXiv:1301.7014
   [hep-ph]].
   
   \bibitem{Gallardo:1996aa}J. C. Gallardo, R.~B. Palmer, A.~V.~Tollestrup,
   A. M.~Sessler, A.~N.~Skrinsky, C.~Ankenbrandt, S.~Geer and J.~Griffin
   \emph{et al.}, 
    eConf C \textbf{960625} (1996) R4.
   
   \bibitem{Ankenbrandt:1999as}C.~M.~Ankenbrandt, M.~Atac, B.~Autin,
   V.~I.~Balbekov, V.~D.~Barger, O.~Benary, J.~S.~Berg and M.~S.~Berger
   \emph{et al.}, 
    Phys.~Rev.~ST Accel.~Beams \textbf{2} (1999) 081001 [physics/9901022].
   
   \bibitem{Alsharoa:2002wu}M.~M.~Alsharoa \emph{et al.} [Muon Collider/Neutrino
   Factory Collaboration], 
    Phys.~Rev.~ST Accel.~Beams \textbf{6} (2003) 081001 [hep-ex/0207031].
    
\end{thebibliography}
   \end{document}